\documentclass[3p, preprint, 12pt]{elsarticle}

\usepackage{graphicx}
\usepackage{amssymb}

\usepackage{lineno}
\usepackage{threeparttable}
\usepackage{amsmath}
\usepackage{mathtools}
\usepackage{amssymb}
\usepackage{newtxtext,newtxmath}  
\usepackage{placeins}
\usepackage{bm}
\usepackage{subcaption}
\usepackage{booktabs}
\usepackage{hyperref}
\usepackage{multirow}
\usepackage[ruled,vlined]{algorithm2e}
\usepackage{enumitem}
\usepackage{makecell}
\usepackage{xcolor}

\journal{}

\begin{document}
\begin{frontmatter}





\title{Event-aware analysis of cross-city visitor flows using large language models and social media data}


\author[label1]{Xiaohan Wang}

\author[label1,label2,label3]{Zhan Zhao\corref{cor1}}

\author[label1]{Ruiyu Wang}

\author[label4,label5]{Yang Xu}

\cortext[cor1]{Corresponding author (zhanzhao@hku.hk)}

\affiliation[label1]{{Department of Urban Planning and Design, The University of Hong Kong, Hong Kong}
           }
           
\affiliation[label2]{{Urban Systems Institute, The University of Hong Kong, Hong Kong}
           }

\affiliation[label3]{{Musketeers Foundation Institute of Data Science, The University of Hong Kong, Hong Kong}
           }
           
\affiliation[label4]{{Department of Land Surveying and Geo-informatics, The Hong Kong Polytechnic University, Hong Kong}
            }

\affiliation[label5]{{Research Centre for Digital Transformation of Tourism, The Hong Kong Polytechnic University, Hong Kong }}

\begin{abstract}
Public events, such as music concerts and fireworks displays, can cause irregular surges in cross-city travel demand, leading to potential overcrowding, travel delays, and public safety concerns. To better anticipate and accommodate such demand surges, it is essential to estimate cross-city visitor flows with awareness of public events. Although prior studies typically focused on the effects of a single mega event or disruptions around a single venue, this study introduces a generalizable framework to analyze visitor flows under diverse and concurrent events. We propose to leverage large language models (LLMs) to extract event features from multi-source online information and massive user-generated content on social media platforms. Specifically, social media popularity metrics are designed to capture the effects of online promotion and word-of-mouth in attracting visitors. An event-aware machine learning model is then adopted to uncover the specific impacts of different event features and ultimately predict visitor flows for upcoming events. Using Hong Kong as a case study, the framework is applied to predict daily flows of mainland Chinese visitors arriving at the city, achieving a testing R-squared of over 85\%. We further investigate the heterogeneous event impacts on visitor numbers across different event types and major travel modes. Both promotional popularity and word-of-mouth popularity are found to be associated with increased visitor flows, but the specific effects vary by the event type. This association is more pronounced among visitors arriving by metro and high-speed rail, while it has less effect on air travelers. The findings can facilitate coordinated measures across government agencies and guide specialized transport policies, such as shuttle transit services to event venues, and comprehensive on-site traffic management strategies.

\end{abstract}




\end{frontmatter}



\section{Introduction}

Understanding cross-city travel demand is crucial for the efficient operation of intercity transport infrastructure, including airports, railway stations, highways, and shuttle transit systems. However, such travel demand often fluctuates irregularly due to various external factors, such as extreme weather conditions \citep{lu_inter-city_2014, yang_impact_2021} and holidays \citep{li_exploring_2020}. In recent years, public events, such as concerts and fireworks displays, have become an increasingly
significant driver of cross-city travel demand surges. Policymakers actively leverage frequent and diverse events to attract visitors and boost local economy \citep{pike_destination_2014}. The combination of widespread event promotion on social media and improved intercity transport connectivity further encourages people to travel across cities to attend a wide range of events \citep{zhao_urban_2019, wang_evolution_2018, zeng_what_2014-1, XIANG2010179, li_exploring_2020}. 
Existing studies examining the impact of public events on travel demand are relatively limited and often focus on a specific event venue using a single data source \citep{chen_subway_2020,LIANG2024102153,pereira_why_2015,rodrigues_bayesian_2018}. Some researchers have considered public events as disruptive factors affecting visitor arrivals \citep{heller_how_2021, zhang_mitigating_2024}, but usually with a sole focus on occasional international mega events or specific sports games. 
The joint effect of diverse and concurrent events on cross-city visitor flows remains a largely unexplored research area.  


It is well recognized that popular events can draw large crowds, leading to unexpected congestion and overcrowding at cross-city transport facilities, local public transit systems, and specific event sites. Timely insights about cross-city visitor flows are critical to facilitate coordinated measures across government agencies and guide specialized transport policies. The operators of cross-city transport facilities can identify key entry points susceptible to these disruptions and adjust the allocation of personnel and resources accordingly to minimize delays. Local public transit operators should offer shuttle services to event venues, while also scaling up operations by increasing service frequency and extending operating hours. At specific event sites, transport authorities should collaborate closely with traffic police, emergency services, parking facility managers, and event organizers to implement comprehensive on-site traffic and crowd management strategies. However, local authorities often lack a systematic approach to estimate potential visitor flows attracted by these diverse events in advance, due to several factors: (1) event information is fragmented across multiple online sources, as events are organized by different entities; (2) while many events are ticketed, visitors do not necessarily book them prior to the event day, and there are certain events, such as fireworks displays, that do not involve ticket sales; and (3) multiple events can occur on the same day, which may lead to synergetic effects on cross-city visitor flows. 
Therefore, to effectively manage such irregular demand surges, it is much needed to develop a systematic event-aware analytical framework for dynamic prediction of cross-city visitor flows under diverse and concurrent events.

In today's digital age, social media data serve as a valuable resource for estimating the potential popularity of events. On the one hand, these platforms play a pivotal role in events marketing \citep{hu_tourism_2022} before events take place. On the other hand, participants often share their experiences on social media after attending events, creating word-of-mouth endorsements \citep{hu_tourism_2022,cui_operational_2018,duan_online_2008,dellarocas_exploring_2007}. Both pre-event promotion and post-event experience sharing can significantly affect individuals' intentions to attend events \citep{li_tourism_2023,hudson_effects_2015}, and 
user engagements with these social media posts can serve as an indicator for the events' potential to attract visitors in the future
\citep{boivin_analysis_2019,hudson_effects_2015,zeng_what_2014-1,XIANG2010179,DEDEOGLU2020103954}. Therefore, social media popularity of events may be incorporated as a predictive factor for visitor flows \citep{zhang_mitigating_2024}.
However, quantifying such popularity requires extracting meaningful and relevant information from noisy data. Previous studies on short-term demand predictions often focused on the number of social media check-ins as an indicator of user attendance within subsequent hours \citep{ni_forecasting_2016,xue_forecasting_2022}. In these cases, relevant posts can be directly filtered using the post times and geo-tags. However, effective transport service adjustments (e.g., special shuttle services) often require visitor flow prediction at least days before the event. Thus it is essential to collect early social media posts that promote the event or share users' experiences related to it, which requires semantic analysis of the social media post contents. 
Existing studies often manually screen promotional posts from official accounts \citep{pereira_why_2015}, overlooking user-generated contents that share personal event experiences. 
To systematically extract social media popularity from both promotional and experience-sharing posts, more efficient and reliable natural language processing techniques are needed.

To address these challenges, this study presents an event-aware analytical framework to extract event features from online and social media data and then utilize these features for cross-city visitor flow prediction. First, we compile comprehensive event data across various categories and scales in a unified data structure. To incorporate social media popularity as an indicator of potential event appeal, we propose to leverage large language models (LLMs) to process multi-source event information online and analyze the semantics of massive user-generated content on social media platforms. A Gradient Boosted Decision Trees (GBDT) rolling prediction model is then adapted to incorporate these event features and predict visitor flows for upcoming events. The proposed framework is demonstrated using Hong Kong as a case study. The specific contributions of this study are summarized as follows:

\begin{itemize}[noitemsep]
  \item A generalizable LLM-based pipeline is developed to extract event features from disparate data structures. LLMs facilitate the mining of event features from online sources and enable the semantic analysis of extensive social media data.
  \item We use social media popularity metrics to estimate the impact of pre-event promotions and post-event word-of-mouth on visitor attraction, with their effectiveness assessed through comparative analysis.
  \item Incorporating a comprehensive set of event features, a GBDT rolling prediction model is then adapted to predict visitor flows for upcoming events at least days in advance, which can facilitate coordinated measures across government agencies and guide specialized transport policies.
  \item Using Hong Kong as a case study, empirical analysis verifies the contribution of event features to improved visitor flow prediction performance. The results highlight heterogeneous effects across event types and major travel modes, offering insights for crowd management and strategic operation of transport services and infrastructure.
  

\end{itemize}

\section{Literature Review}

\subsection{Special Events and Cross-City Travel}

Cross-city travel demand modeling has traditionally focused on macro-level human mobility and regular commuting demand within metropolitan areas. Researchers have employed gravity models to analyze cross-city travel patterns, considering factors such as population size, land use, and travel behaviors \citep{zhao_revisiting_2023,wirasinghe_aggregate_1998}. Additionally, complex network analysis has been utilized to uncover the structure of cross-city commuting systems, providing insights into mobility flows and spatial interactions \citep{de_montis_modeling_2010,de_montis_structure_2007}.
However, these research primarily emphasizes regular travel patterns while often overlooking scenarios where demand experiences sudden, significant fluctuations due to external disruptions. 

The demand for cross-city travel often exhibits irregular fluctuations influenced by various factors, such as pandemics \citep{wei_spatiotemporal_2024,wang_analysis_2022,yamaguchi_pattern_2023}, extreme weather conditions \citep{lu_inter-city_2014,yang_impact_2021}, holidays \citep{li_exploring_2020}, etc. In recent years, public
events, such as concerts and fireworks displays, have become an increasingly significant driver
of cross-city travel demand surges. Policymakers leverage frequent and diverse events to attract visitors and boost local economy \citep{pike_destination_2014}. The continuous advancement of metropolitan transportation systems has further facilitated people to travel across cities more conveniently \citep{wang_evolution_2018,li_exploring_2020}. As a result, individuals are more inclined to undertake cross-city trips to participate in various events. Some popular events unexpectedly attract large crowds, placing significant strain on cross-city transportation infrastructure and public transit systems.
Given the growing role of events as a motivator for cross-city travel, their impact should be integrated into the analysis of cross-city visitor flows.

\subsection{Impact of Public Events on Travel Demand}

Previous research on the impact of public events on travel demand has largely concentrated on events occurring at a single venue. These studies aim to decompose the influence of such events and predict travel demand during event periods \citep{pereira_why_2015, rodrigues_bayesian_2018, chen_subway_2020, LIANG2024102153}. For instance, an ARIMA model has been applied to predict metro passenger flows surrounding the Nanjing Olympic Sports Center over several event days \citep{chen_subway_2020}. To evaluate the effect of events around a large venue, researchers decomposed the impact of these events alongside regular commuting patterns, and their models were tested using metro ridership data in Singapore \citep{pereira_why_2015, rodrigues_bayesian_2018}. 

In the field of tourism demand forecasting, the concept of event-motivated cross-city travel dates back to the nineteenth century \citep{pike_destination_2014}. However, fewer studies have examined public events as a factor that disrupts visitor flows in demand predictions \citep{heller_how_2021,zhang_mitigating_2024}. For example, \cite{zhang_mitigating_2024} integrated disturbances from both crises and mega-events into tourism demand forecasting for two cities, using online search trends related to events as predictive variables. Nevertheless, this study was limited to examining four infrequent international mega-events. Today, policymakers have increasingly relied on more frequent and diverse events to attract visitors and maintain urban vibrancy, such as music concerts and exhibitions. The joint impact of multiple concurrent events on city-wide visitor flows remains a largely unexplored area of research.

\subsection{Opportunities and Challenges in Integrating Social Media Data}

Events can vary significantly in scale, from large music concerts to small exhibitions. Estimating attendance is particularly challenging for events held in open spaces or those not requiring reservations. To quantify an event's potential to attract crowds and incorporate it as a measurable feature, prior studies have employed topic modeling techniques or more recent approaches such as LLMs to interpret and categorize event topics from textual data \citep{rodrigues_bayesian_2018, LIANG2024102153}. However, analyzing social media engagement data provides a more direct measure of an event's popularity compared to relying solely on textual descriptions \citep{pereira_why_2015,zhang_mitigating_2024}.

In recent years, social media platforms such as \textit{Facebook} and \textit{Instagram} have become essential marketing channels for event organizers and marketing organizations \citep{zhou_tourism_2024}. 
Events are typically one-off or periodic, resulting in fewer information sources available for potential visitors \citep{hu_tourism_2022}. When planning trips motivated by events, users often turn to these social media channels for event-related information. 
Social media data encompasses not only event advertisements from event marketing organizations but, more importantly, user-generated content that shares participants' personal experiences. Studies indicate that social media information affects users' behavior through two mechanisms: the attention effect and the endorsement effect \citep{hu_tourism_2022,cui_operational_2018,duan_online_2008,dellarocas_exploring_2007}. 
Promotional posts, created by event organizers or marketers, enhance the events' online exposure, contributing to the attention effect. User-generated content, which shares attendees' personal experiences, serves as an endorsement and fosters online word-of-mouth communication. Driven by both online promotional efforts and extensive word-of-mouth communication, these posts jointly impact people's intentions to travel across cities to attend the events \citep{boivin_analysis_2019,hudson_effects_2015,zeng_what_2014-1,XIANG2010179,DEDEOGLU2020103954}.


However, social media data are noisy and often involve irrelevant posts. Table~\ref{tab:literature} summarizes studies that integrate social media information to predict travel demand during events and outlines their methods for identifying event-related posts. For short-term prediction of travel demand (e.g., next-hour prediction), researchers usually identify event-related posts based on spatiotemporal proximity to capture real-time attendance behaviors \citep{ni_forecasting_2016, xue_forecasting_2022}. These studies primarily focus on social media check-in records that disclose users' real-time locations, reflecting their event-going behaviors. 
Specifically, this approach first selects posts generated a few hours before the event start time around the event venue, and then incorporates features such as the number of posts and unique post creators into prediction models \citep{ni_forecasting_2016, xue_forecasting_2022}. However, it relies mainly on the spatiotemporal attributes of social media posts, with less emphasis on the semantic relevance of the posts.

In contrast, to effectively plan and manage the sudden surge of travel demand due to events, it is necessary to predict visitor flows at least days in advance. In this case, the focus shifts from social media check-ins to posts that promote the event or share personal experiences, as they influence other people's intentions to attend. Since these promotional and experience-sharing posts are not necessarily posted at the event venue or within the event's time frame, it is no longer sensible to filter social media posts based on spatiotemporal relevance. Instead, semantic relevance becomes more important, which requires new methods to comprehend the content of social media posts rather than their time or location. Although previous studies have used variables such as \textit{Facebook} ``likes" and \textit{Google} hits to explain metro ridership during events \citep{pereira_why_2015}, they still largely relied on manual filtering and only considered posts from marketing organizations' official event pages or accounts, overlooking the significant influence of user-generated content. This is problematic as people's perceptions of events are shaped not only by official marketing but also by word-of-mouth communication on social media \citep{hudson_effects_2015,lei_organic_2023}. The reliance on manual screening further limits the method's scalability. Therefore, there is a pressing need for more efficient and reliable data processing methods to enhance semantic analysis for large-scale user-generated content.

\begin{table}[!ht]   
    \caption{Incorporation of social media data in studies on event-related mobility}\label{tab:literature}
    \centering  \footnotesize
    \begin{tabular}{p{2cm}p{3.5cm}p{2cm}p{3.5cm}p{3cm}}
        \toprule
        Authors & Topics & Data Sources & Relevance Identification & Feature Extraction \\
        \midrule
        \citet{ni_forecasting_2016}& Real-time traffic event detection and passenger flow predictions within the following hour & Twitter & Geo-tags, time window, hashtags & Number of event-related posts/ unique tweet users \\
        \midrule
        \citet{xue_forecasting_2022} & Passenger flow predictions under events within the following hour & Sina Weibo &  Geo-tags, time window & Number of event posts \\
        \midrule
        
        \citet{pereira_why_2015} & Event-related crowding explanation&Facebook, Google & Search through geo-tags and hashtags, and filter manually & Event topics, categories, \textit{Facebook} ``likes", \textit{Google} hits
        \\
              \midrule
        \citet{du_predicting_2014} & Event Attendance prediction in Event-Based Social Networks & Douban & - & User interest with a forgetting function\\
                \bottomrule
    \end{tabular}
\end{table}

In recent years, the field of natural language processing has experienced a significant transformation with the emergence of LLMs, such as the Generative Pre-trained Transformer (GPT) series \citep{Achiam2023GPT4TR}. LLMs demonstrate exceptional abilities in understanding and generating human-like text. In relevant fields, \citet{LIANG2024102153} highlighted the proficiency of LLMs in summarizing event descriptions, demonstrating their performance to be comparable to humans for this specific task. In addition, LLMs have been utilized in tasks such as understanding urban concepts \citep{fu_towards_2024} and identifying location descriptions \citep{mai_opportunities_2024}. Given these capabilities, LLMs hold substantial potential to address the challenges in event dataset construction and feature extraction, such as aiding in extracting event features from textual data, formatting consistent feature structure from multiple online sources, understanding the semantics of social media posts and event descriptions, etc. Leveraging LLMs will potentially enhance the efficiency and applicability of large-scale social media data analysis, which could provide valuable insights into visitor behavior and enhance the performance of visitor flow prediction \citep{song_tourism_2024}.

\section{Study Area and Datasets}
\subsection{Study Area}
\begin{figure}[!ht]
  \centering
  \includegraphics[width=1\textwidth]{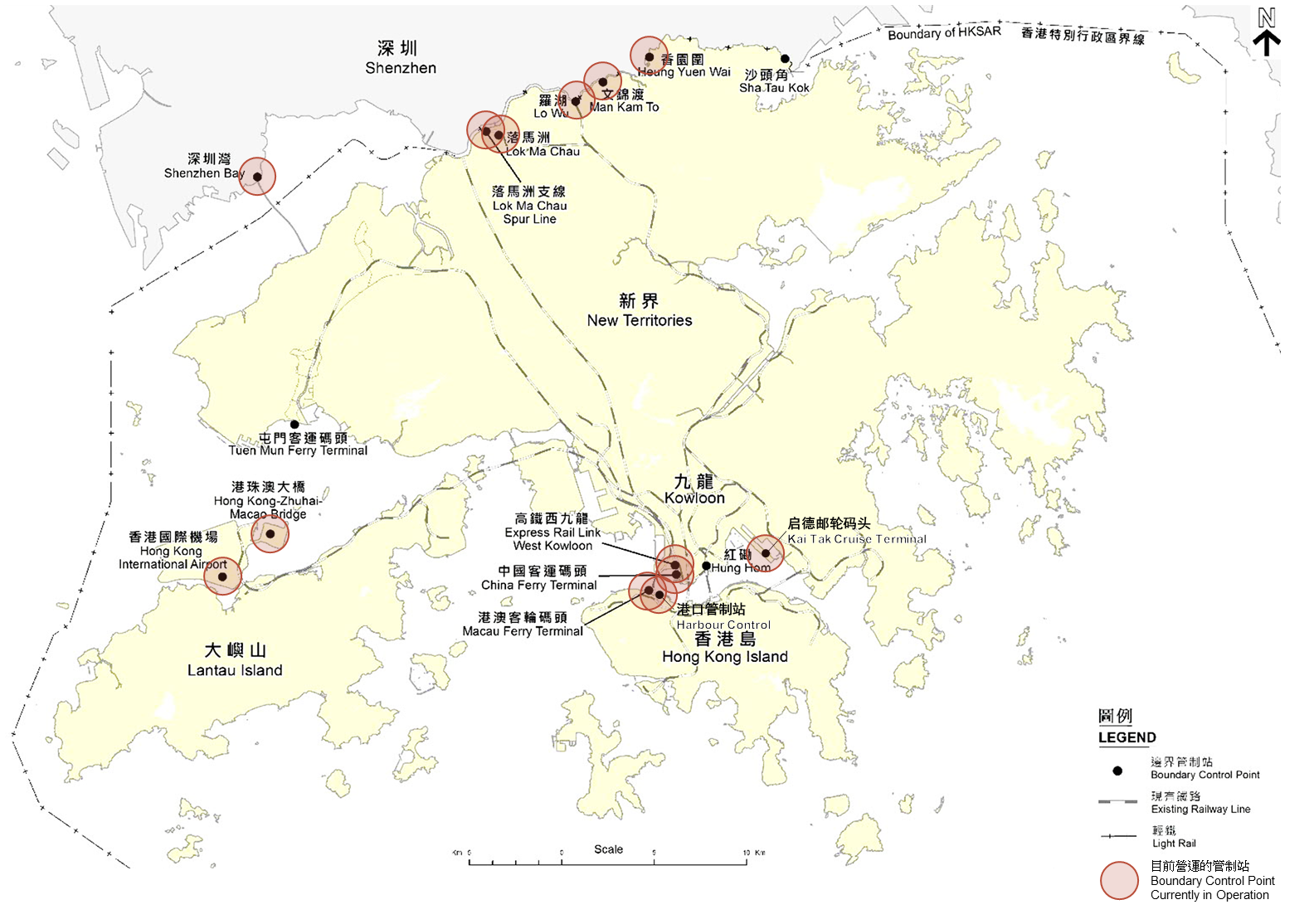}
  \caption{Cross-border entry points in Hong Kong 
  (adapted from \citet{survey2021})
  }\label{fig:map}
\end{figure}

We choose Hong Kong as the study area, due to its strategic policy focus on hosting diverse public events to attract visitors and drive economic growth. 
Located in southern China within the Greater Bay Area (GBA), Hong Kong maintains close geographical ties and experiences frequent flows of people and economic activities with mainland China. 
As shown in Fig.~\ref{fig:map}, Hong Kong currently operates four major land customs entry points and two metro entry points along the border with Shenzhen, one of the biggest cities in China. Additionally, a cross-sea bridge checkpoint connects Hong Kong with Macau and Zhuhai; four ferry entry points link cities in the region. West Kowloon High-Speed Rail (HSR) Station and Hong Kong International Airport provide direct links to numerous mainland Chinese cities. These transportation hubs serve as primary channels for cross-border visitors from mainland China.

Hong Kong has been a popular destination for visitors from mainland China, especially those coming from neighboring cities. In 2017, mainland Chinese visitors contributed to an average daily arrival rate of 216,600 in Hong Kong, with 57.8\% being same-day trips and 91.8\% of these visitors originating from GBA (excluding Hong Kong residents) \citep{survey2017}. However, the outbreak of the COVID-19 pandemic severely disrupted leisure travel by mainland Chinese visitors \citep{survey2017,survey2021,tsui_analyzing_2021}. Following the reopening of customs in January 2023, Hong Kong has strategically hosted events to attract visitors and increase tourism revenue \citep{getz_progress_2016,chang_exploring_2022}. These events have attracted substantial numbers of mainland Chinese visitors, putting pressure on customs processing and accommodation capacities. As transportation services are typically designed for regular commuting traffic, the sudden surge of visitor arrivals can lead to overcrowding and travel delays, negatively affecting the visitors' experience. Predicting visitor flows for upcoming events is thus essential for managing visitor demand, as it allows for proactive adjustments, such as assigning additional frontline staff at key entry points, enhancing local public transit service capacities, and deploying traffic management measures around event sites. 

\subsection{Datasets}
\subsubsection{Visitor Flows, Holiday and Weather Data}
\begin{figure}[!ht]
  \centering
  \includegraphics[width=1\textwidth]{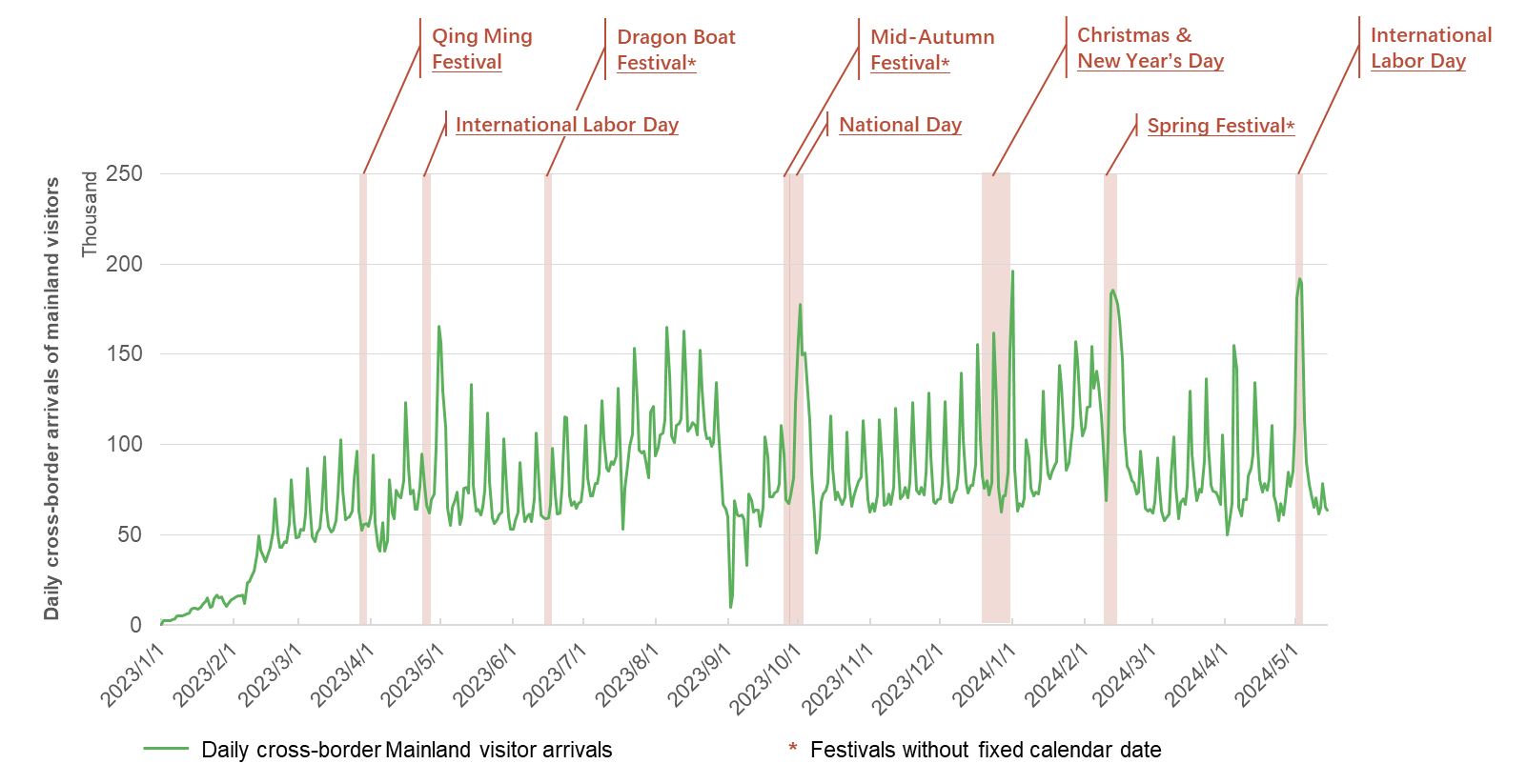}
  \caption{Daily flows (arrivals) of mainland Chinese visitors to Hong Kong from January 2023 to May 2024}\label{fig1}
\end{figure}

Since cross-border arrivals from mainland China resumed in March 2023, we collected data from March 1, 2023, to May 15, 2024. The cross-border visitor arrival data is provided by the Hong Kong Immigration Department\footnote{https://data.gov.hk/en-data/dataset/hk-immd-set5-statistics-daily-passenger-traffic}. Fig.~\ref{fig1} illustrates daily cross-border mainland Chinese visitor arrivals from January 2023 to May 2024, showing the resurgence since the borders reopening.
visitor arrivals exhibit noticeable weekly peaks and sharp increases during calendar events (labeled in red). Additionally, January-February, and July-August are usually the summer and winter school vacation periods in China, respectively. During these periods, visitor arrivals on weekdays experience substantial increases. Given that weather conditions are considered to significantly influence event attendance and travel behavior \citep{zhang_exploring_2021,becken_measuring_2013}, we further obtain daily total rainfall\footnote{https://data.gov.hk/en-data/dataset/hk-hko-rss-daily-total-rainfall}, maximum daily temperature\footnote{https://data.gov.hk/en-data/dataset/hk-hko-rss-daily-temperature-info-hko}, and historical tropical cyclone warning signals (typhoon signals) from the Hong Kong Observatory\footnote{https://www.hko.gov.hk/tc/wxinfo/climat/warndb/warndb1.shtml}. 

\subsubsection{Event Data}

Event data is collected through multiple sources in various formats. We compile event data for Hong Kong starting from March 2023, utilizing a dedicated event website, \textit{Timable}\footnote{https://timable.com}. This platform provides comprehensive historical event information, including concerts, festivals, exhibitions, competitions, performances, etc. The local government also displays selected events on the official tourism and destination branding websites in different structures. As a supplement, we obtain the ``year-round event data'' in 2023 from the \textit{Hong Kong Tourism Board} website\footnote{https://www.discoverhongkong.com/hk-eng/what-s-new/events.html}, the ``mega events list'' in 2024 provided by \textit{Hong Kong Asia's World City} website\footnote{https://www.brandhk.gov.hk/en/mega-events/mega-events}, and the ``major sports list'' from the \textit{Major Sports Events Committee} website\footnote{https://www.mevents.org.hk/en/calendar.php}.

\begin{figure}[!ht]
  \centering
  \includegraphics[width=0.8\textwidth]{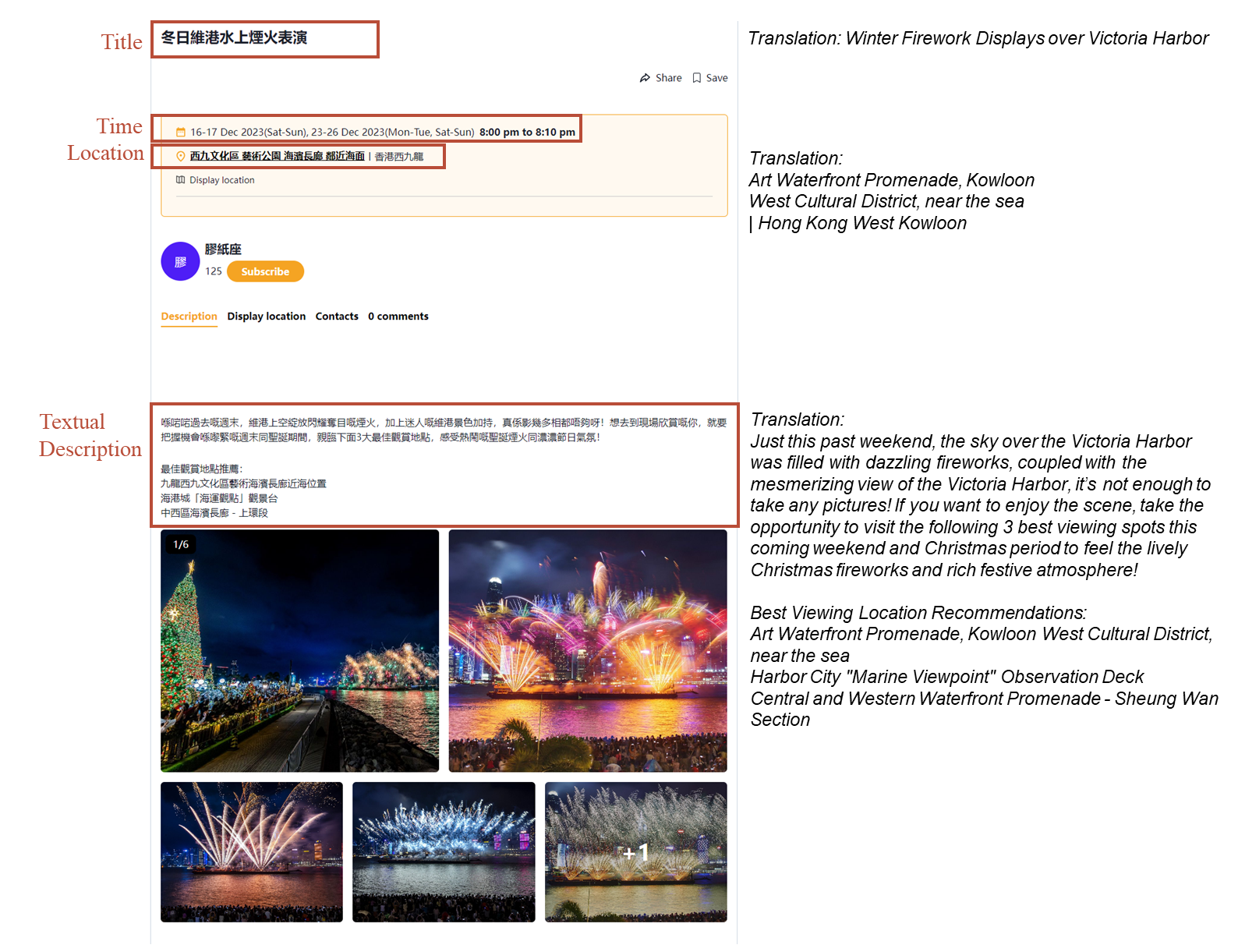}
  \caption{Illustration of the event page of Winter fireworks display on Timable}\label{fig:timable}
\end{figure}

Fig.~\ref{fig:timable} illustrates the event page of a fireworks display on Timable, providing details such as event title, time, location, and description. However, the raw textual event features extracted from web pages are often unstructured, which poses challenges for conventional natural language models to interpret the underlying semantics and generate specific event features. For instance, many events occur over multiple sessions, and conventional models find it difficult to extract structured times for each session from expressions like ``16-17 Dec 2023 (Sat-Sun), 23-26 Dec 2023 (Mon-Tue, Sat-Sun) 8:00 pm - 8:10 pm'', as shown in Fig.~\ref{fig:timable}. Furthermore, event information extracted from government websites is in various formats. To process textual information from diverse sources, we employ LLMs to extract unified key features such as the event title, time, location, and description for each sub-session. The details will be described in Section~\ref{sec:LLMs}.

\subsubsection{Social Media Data}

An event's popularity on social media platforms is a powerful indicator of its potential to attract visitors. Among the popular Chinese social media platforms, \textit{Rednote} (also known as \textit{Xiaohongshu}) has emerged as a prominent player in recent years, boasting over 350 million users. Unlike other online platforms that focus on one-way content generation for destination recommendations, such as \textit{TripAdvisor}, or the specialized event-based social networks, such as \textit{Douban}, \textit{Rednote} encourages vibrant peer communication through a wealth of user-generated content.
\textit{Rednote} users typically share information or personal experiences about the latest trending events, making it a popular tool for discovering event information \citep{lei_organic_2023}. 

\begin{figure}[!ht]
  \centering
  \includegraphics[width=0.9\textwidth]{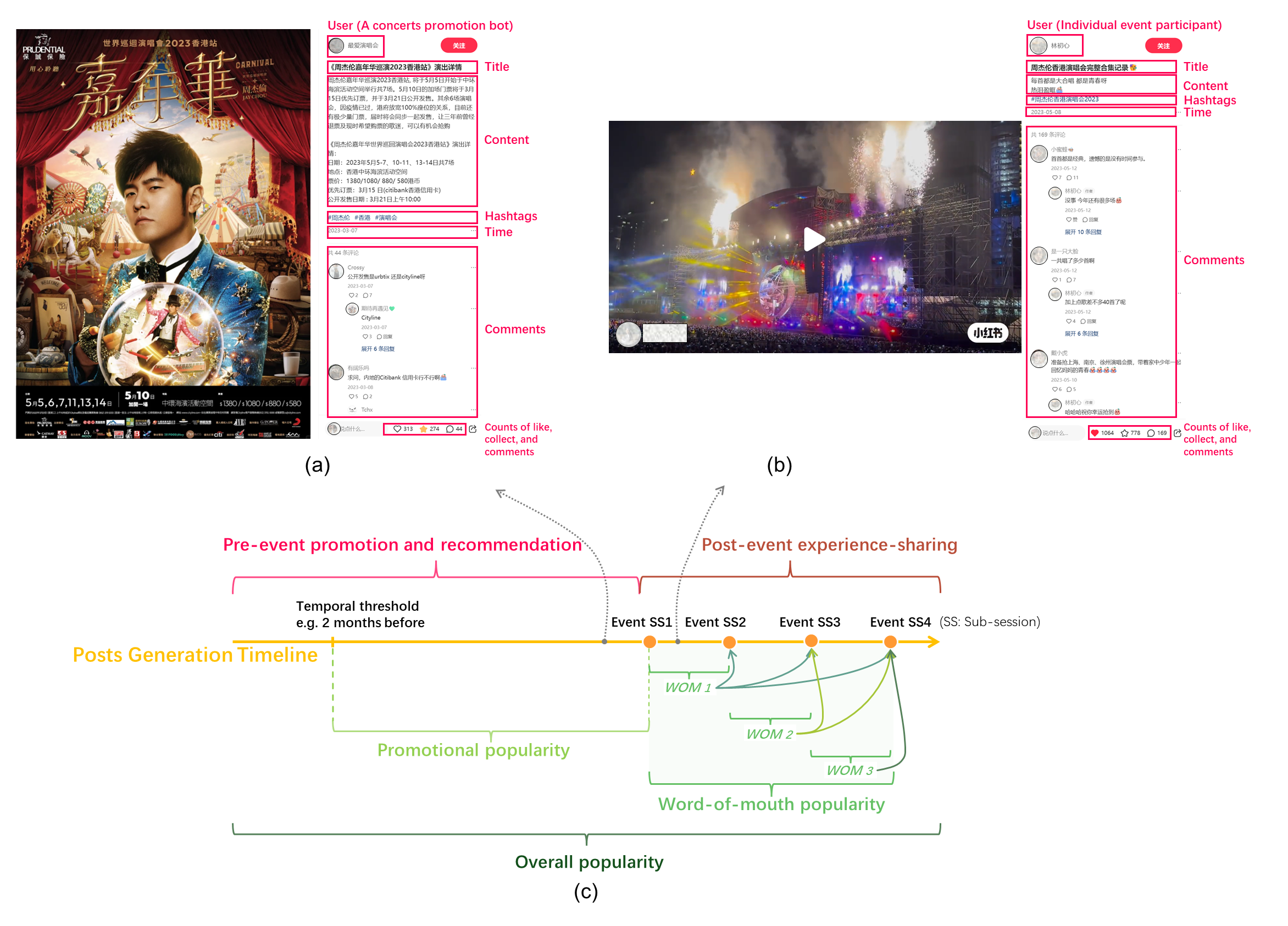}
  \caption{Illustrations of social media posts and popularity measures definition}\label{fig:popularity}
\end{figure}

To assess an event's popularity, we summarize keywords for the events and search for \textit{Rednote} posts using these keywords. For each event, we crawl up to top $g$ posts that have received the most ``likes'', where $g$ is a number to be specified. In this case, without loss of generality, we set $g=100$.
As shown in Fig.~\ref{fig:popularity} (a-b), each post on \textit{Rednote} includes information such as user ID, title, textual content, hashtags, time, and counts of ``like", ``collect", and ``comment".
When \textit{Rednote} users seek information about events, they often click the ``like" or ``collect" buttons to express their interest and bookmark the posts. 
We assume that the total count of ``likes'' and ``collects" for an event's related posts reflects the event's exposure on social media, thereby indicating its potential to attract attendees. In this study, we measure an event's social media popularity by the total number of ``likes" and ``collects" under its posts on \textit{Rednote}. Due to the limitations of search algorithms and keyword ambiguity, irrelevant posts are occasionally retrieved. We will introduce our LLM-enhanced method for processing social media data in Section~\ref{sec:LLMs}. On most social media platforms, users typically express their interests through ``likes". Therefore, we use ``likes" as a general term to refer to user engagement with social media posts. It is worth highlighting that \textit{Rednote} is only used here as a representative social media platform, and the methods introduced in later sections should be easily adaptable to other platforms. 

\section{Methods}

\subsection{LLM-enhanced Event Feature Mining}
\label{sec:LLMs}
\begin{figure}[!ht]
  \centering
  \includegraphics[width=1\textwidth]{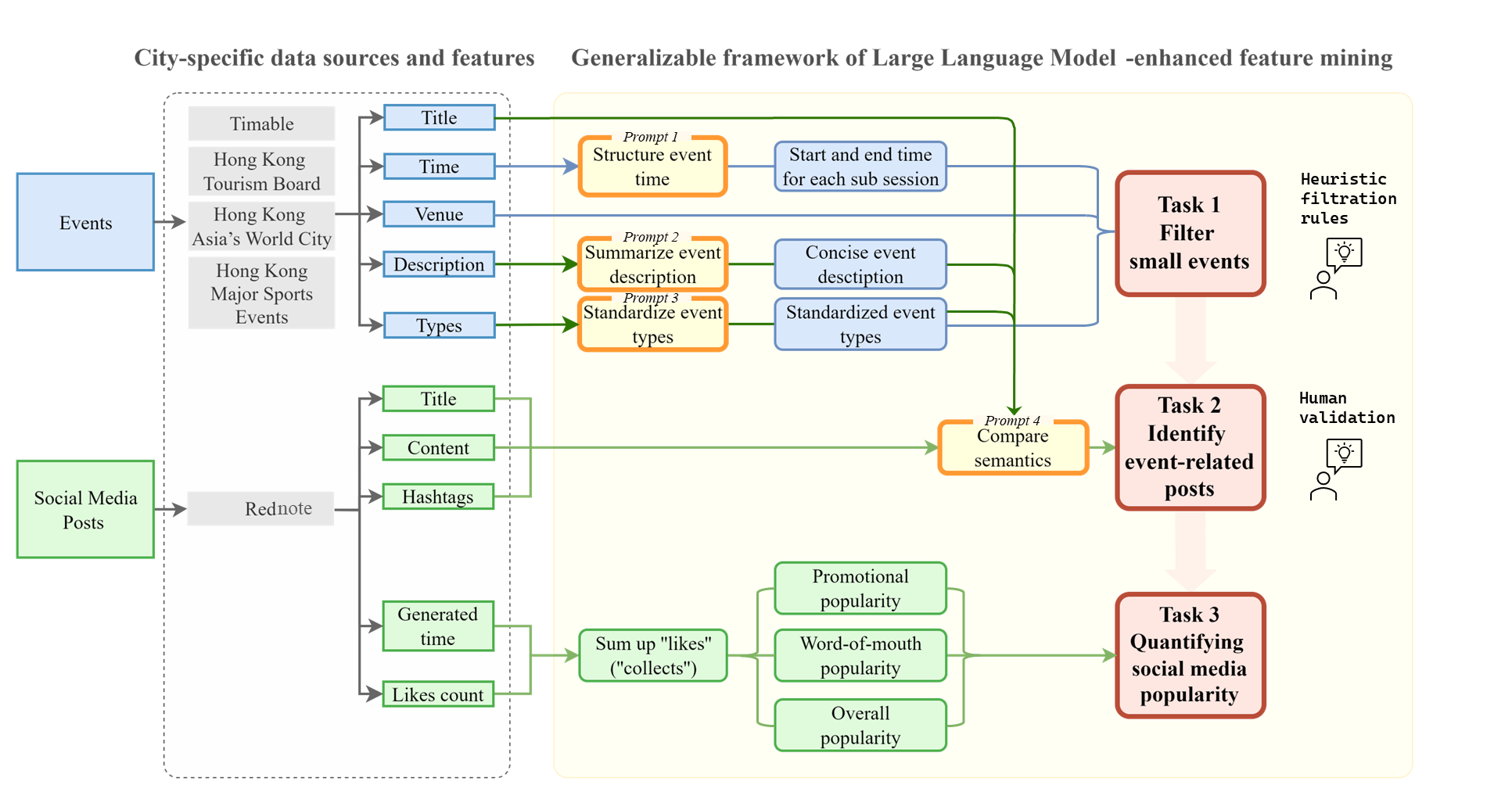}
  \caption{A LLM-enhanced framework of event data collection and feature mining}\label{fig:framework}
\end{figure}

To effectively structure event features from various online data sources, we develop a generalizable pipeline for LLM-enhanced event feature mining. As shown in Fig.~\ref{fig:framework}, this pipeline leverages the advanced text processing and semantic analysis capabilities of LLMs, combined with human expertise, to tackle three main objectives: (1) filtering events, (2) matching social media posts related to events, and (3) quantifying event popularity on social media. 

To filter events, we use LLMs to structure event time, summarize event descriptions, and standardize classifications across multiple data sources. Specifically, we establish heuristics to exclude events that are unlikely to draw visitors from other cities. Smaller events are typically long-term, held at smaller venues, and lack broad public appeal. The details will be described in Section~\ref{sec:Feature Enhancement and Event Filtering}. 

To identify event-related social media posts, we characterize posts based on their titles, content, hashtags, timestamps, and the number of ``likes" they receive. LLMs can be used to compare the semantics of the texts in event descriptions and social media posts, allowing us to pinpoint posts associated with specific events and filter out irrelevant content. The details will be described in Section~\ref{sec:Event-related Social Media Post Identification}. 

To measure social media popularity, we propose three metrics based on the post's timing and function: promotional popularity, word-of-mouth popularity, and overall popularity. Specifically, we distinguish between social media posts that promote events beforehand and those that share experiences afterward; the former are associated with to promotional popularity and the latter word-of-mouth popularity. Overall popularity encompasses all forms of social media engagement with the events, regardless of timing. The effectiveness of these popularity metrics will be evaluated in Section~\ref{sec:Social Media Popularity}. 

\subsubsection{Feature Enhancement and Event Filtering}
\label{sec:Feature Enhancement and Event Filtering}
Event time information is often presented as unstructured text, making it challenging to characterize. To address this, we leverage the ability of LLMs to understand semantics and convert raw event information into structured features. As shown in Table~\ref{prompt}, we design \textit{Prompt 1} to process raw time information by extracting and formatting specific start and end time for each session. \textit{Prompt 2} directs the LLM to summarize event descriptions while omitting details about timing, location, ticketing, and promotion. To ensure these summaries are both comprehensive and concise, we impose limits on the LLM's token usage during summary generation. These prompt templates can be adapted for specific events by modifying placeholders as needed.

Different types of events vary in their appeal to visitors, requiring consistent classification rules across various data sources. To achieve this, we predetermine the scope of event types based on existing classifications from specialized event websites and local tourism portals. As shown in \textit{Prompt 3}, the LLM categorizes events into these predefined types, standardizing event classification.

To exclude smaller events that are less likely to attract visitors from other cities, we establish three heuristic rules. First, we limit the number of sub-sessions to exclude long-term events, such as individual art exhibitions, which typically draw local visitors continuously and are unlikely to attract a significant number of cross-city visitors in the short term. Second, we exclude events held at small venues, such as ground floors of shopping malls, by compiling a list of large-capacity venues commonly used for each event type and filtering events accordingly. Third, we focus on certain event types, including concerts, exhibitions, sports competitions, and fireworks displays, as small-scale events like weekend fairs are less likely to be promoted to visitors. These rules help us focus on events with the potential to influence visitor flows, allowing for more efficient collection and processing of social media data.

\begin{table} 
\caption{LLM prompt templates for event feature mining}\label{prompt}
\centering \footnotesize
\begin{tabular}{p{1.2cm}p{2cm}p{12cm}}
\toprule
  \textbf{ID} & \textbf{Task} & \textbf{Prompt template}  \\
  \midrule
  Prompt 1 & Structuring event period &
    {Below is the raw description of an event's hosting time, including event ID: \textit{\{event ID\}}, and event time: \textit{\{raw time description\}}.
    
    Your task is to extract the exact start and end datetime for each event. If an event has multiple sessions, find the exact start and end datetime for each session. If a session does not have a specific start and end time, set the start time to ``00:00:00''and the end time to ``23:59:59'' of the event day.
    
    Provide the output in four columns:
    \begin{itemize}[noitemsep]
        \item Id: The event ID.
        \item Sub id: A unique number for each session of the event, starting from 1. 
        \item Start time: The start datetime of the event session in ``YYYY-MM-DD hh:mm:ss'' format.
        \item End time: The end datetime of the event session in ``YYYY-MM-DD hh:mm:ss'' format.
    \end{itemize}
    }\\
  \midrule
  Prompt 2 & Summarizing event descriptions &
    {Please summarize the event information given the raw event description extracted from the web page: \textit{\{raw event description\}}.
    
    Requirements are as follows:
    \begin{itemize}[noitemsep]
        \item 
        Exclude details about time, location, registration, tickets, and payment; only focus on the event content.
        \item 
        Be concise within \textit{\{token usage\}} tokens.
        \item 
        Summarize in \textit{\{language\}}.
        \item 
        Only provide the summarized event description without any additional explanations.
    \end{itemize}
    } \\
    \midrule

  Prompt 3 & Standardizing event classification & 
    {Below is a description of an event that occurred in \textit{\{study area\}}, called \textit{\{event title\}}. \textit{\{summarized event description\}}. Please match the events into the following categories:  
    \textit{\{A list of predefined event types\}}. 
    
    Only provide the event type without any additional description.
    } \\
    
    \midrule
  Prompt 4 & Identifying event-related social media posts & 
    {Below, you will find a description of an event and the content of a social media post. Please compare their semantics and determine if the post {post ID} is related to the given event. Respond with ``Yes" if the post is related, or ``No" if it is not.
    
    Event Information:
        \begin{itemize}[noitemsep]
        \scriptsize
        \item
        Title: \{\textit{event title}\}
        \item
        Type: \{\textit{event type}\}
        \item
        Description: \{\textit{summarized event description}\}
        \end{itemize}
    
    Social Media Post Details:
        \begin{itemize}[noitemsep]
        \scriptsize
        \item
        Title: \{\textit{post title}\}
        \item
        Content: \{\textit{post content}\}
        \item
        Geo-tags: \{\textit{post geo-tags}\}
        \item
        Hashtags: \{\textit{post hashtags}\}
    \end{itemize}
    
    Criteria for Relevance:
        \begin{itemize}[noitemsep]
        \item
        The post must reference an event occurring in \{\textit{study area}\}.
        \item
        The post's content must demonstrate a clear connection to the event based on the provided event information.
        \end{itemize}
    } 
\\
  \bottomrule
\end{tabular}
\end{table}

\subsubsection{Event-related Social Media Post Identification}
\label{sec:Event-related Social Media Post Identification}
Due to the ambiguity of search algorithms in social media platforms, irrelevant information would be captured when using keyword searches. To accurately identify social media posts related to public events, we use LLMs to compare the semantics of event titles, types, and summarized descriptions with the information found in social media posts, including titles, content, geo-tags, and hashtags. As illustrated in Table~\ref{prompt}, \textit{Prompt 4} assists in determining whether a post pertains to an event taking place in the city within a specific time frame.

To evaluate the reliability and accuracy of our LLM-based method, we compare it against the manual screening method used in prior studies. To do this, we utilize a stratified sampling approach by randomly selecting 5\% of events from each event category, and then manually assess the relevance of social media posts related to these events. The results will be compared with those produced by the LLM-based approach. Our validation indicates that the LLM aligns with human judgments approximately 90\% of the time. Discrepancies generally arise in complex contexts that also challenge human evaluators, such as posts recommending multiple events or discussing the outcomes of specific sports games. Overall, LLMs prove to be highly efficient in identifying event-related posts from large datasets, and the results consistently align with those produced by human evaluators.

\subsubsection{Social Media Popularity}
\label{sec:Social Media Popularity}

To measure an event's popularity on social media, we distinguish between overall, promotional, and word-of-mouth popularity. Overall popularity is determined by the total number of ``likes" on all posts related to the event regardless of post time. We further classify posts into two categories based on their timing: pre-event promotional posts and post-event experience-sharing posts. To ensure the posts are pertinent to the current event, we establish a temporal threshold to exclude posts from previous years or seasons. The ``likes” on pre-event promotional posts indicate an event's popularity before it happens and reflect the effectiveness of promotion efforts, which we term promotional popularity.

After each event session, attendees frequently share their experiences on social media, creating a word-of-mouth effect that can influence others' decision to attend future sessions. In events with multiple sessions, posts about an early session can affect attendance at later sessions.  As illustrated in Fig.~\ref{fig:popularity} (c), $WOM_1$ represents the word-of-mouth effect from posts created after sub-session 1 and before sub-session 2. This effect can influence attendance in sub-sessions 2, 3, and 4. This pattern continues with subsequent sessions. Based on this mechanism of online word-of-mouth communication, we define word-of-mouth popularity ($WOMP$) for session $n$ in an event with $N$ total sessions as follows: 
\begin{equation}
WOMP_n=\sum_1^{n'} WOM_{n'}/(N-n')
\end{equation}
where $n' \in \{1, ..., N \}$ is the session number. The word-of-mouth effect is not applicable to the first session or single-session events; therefore, we calculate the word-of-mouth popularity ($WOMP$) starting from session $n = n'+1$. $WOM_{n'}$ represents the word-of-mouth generated between sub-session $n'$ and the subsequent sub-session $n$. It is calculated by summing the ``likes" on social media posts within this period. Posts that share personal experiences from sub-session $n'$ are likely to be recommended to users interested in the event. These users could potentially attend any of the subsequent sub-sessions, based on their availability. Given this, we assume this word-of-mouth effect, $WOM_{n'}$, is evenly distributed across the subsequent sessions, and thus, we divide $WOM_{n'}$ by the number of sessions remaining after sub-session $n'$, which is $N-n'$.


To evaluate the effects of promotional popularity and word-of-mouth popularity on visitor flow prediction, we will test a plethora of model specifications, as detailed in Table~\ref{tab:vars}.

\subsubsection{Parameter Setting}

The parameter setting in the LLM-enhanced event feature mining process for this study is specified below: 
\begin{itemize}[noitemsep]
    \item
    The LLM, \textit{\{GPT 4\}}, is used to enhance event filtration and feature extraction processes \citep{Achiam2023GPT4TR}. Our study focuses on events held in \textit{\{Hong Kong\}} during \textit{\{year of 2023 and 2024\}}. 
    \item
    For summarizing event descriptions in \textit{Prompt 2}, we limit the token usage to \textit{\{120\}} per event. Following the original language used in online sources, the summaries are produced in \textit{\{Chinese\}}.
    \item
    In \textit{Prompt 3}, we define a list of event types based on the existing classifications from dedicated event websites and local tourism websites: \textit{Timable, Hong Kong Tourism Board, Hong Kong Asia's World City}, and \textit{Hong Kong Major Sports Events}. The predefined event types include \textit{\{music concerts, exhibitions, sports competitions, fireworks displays, fairs, performances, and religious activities\}}.
    \item
    In filtering small events, we keep the event types in \textit{\{music concerts, exhibitions, sports competitions, and fireworks displays\}}. The maximum number of sub-sessions is set to \textit{\{30\}}, assuming events with more than 30 sub-sessions are less likely to cause significant travel demand surges. 
    \item
    To quantify social media popularity, we apply a temporal threshold, filtering out posts created more than \textit{\{two months\}} before the event.
\end{itemize}


\subsection{Key Variables}

To understand the factors influencing visitor flows, our model includes variables such as holiday features, weather conditions, and temporal trends. To explore how different types of events attract visitors, we also include the event's social media popularity features by type. The explanatory variables are summarized in Table~\ref{tab:vars}.

\subsubsection{Holiday Features}

Visitor flows exhibit clear weekly cyclical patterns, along with significant peaks during major holidays. To capture the day-of-week effect, we include six binary variables to indicate each day of the week, excluding Tuesday (omitted as a reference category). Holiday effects are determined based on the primary origin of visitors, which, in this context, is mainland China. The major holidays considered in the study include the Spring Festival, Qingming Festival, International Labor Day, Dragon Boat Festival, Mid-autumn Festival, National Day, Christmas, and New Year’s Day. Typically, longer holidays attract more visitors, with more visitor flows at the beginning of the holiday period compared to the end. To incorporate the influence of holiday length, we calculate the number of days remaining until the holiday concludes (e.g., a value of 3 signifies three days remaining, including the present day, whereas 0 denotes a standard weekday). Additionally, major holidays can cause fluctuations outside the usual patterns, prompting the inclusion of two binary variables: one indicating the day before a holiday, and another identifying the week surrounding a holiday, either before it starts or after it ends. We also account for the proximity to holidays by including a variable that tracks the number of days to the nearest holiday, whether previous or upcoming.

Peak tourism seasons often align with school holidays, which we capture using a binary variable labeled school holidays. This reflects special periods in the academic calendar of primary and secondary schools. Specifically for the GBA areas during the study period, the summer vacation spans from July 9, 2023, to August 31, 2023, and the winter vacation stretches from January 22, 2024, to February 19, 2024. Including these variables allows the model to accommodate the impact of both regular and seasonal variations in visitor flows.

\subsubsection{Weather Features}

Weather conditions significantly influence travel behavior. We include variables for daily rainfall measured in millimeters, daily maximum temperature measured in Celsius, and typhoon days when a tropical cyclone warning signal is issued. In total, three variables are included in the analysis.

\subsubsection{Temporal Trends}

Temporal trends provide crucial indicators of seasonal fluctuations in visitor flows. To capture such temporal trends, especially in situations with limited historical data, we introduce the Weighted Moving Average (WMA) changing rate as a feature. The WMA rate smooths the time series data by assigning progressively smaller weights to older observations, thereby emphasizing recent data points. This approach helps mitigate the impact of short-term fluctuations and highlights the underlying trend.

The MWA $M$ for visitor flow on day $t$ over the past $P$ days is calculated using the formula:
\begin{equation}
M_t = \frac{\sum_{p=0}^{P-1} \mu_p \ast x_{t-p}}{\sum_{p=0}^{P-1} \mu_p}
\end{equation}
where $x_{t-p}$ represents the visitor flow on day $t-p$, and $\mu_p$ denotes the weight for day $t-p$. In our study, we use a linear decreasing weight $\mu_p = P-p$. 

To predict the visitor flows on the day $t$, we use the changing rate of the MWA of the day $t-1$, which captures the up-to-date trend, defined as:
\begin{equation}
C_t = \frac{M_{t-1} - M_{t-2}}{M_{t-2}}
\end{equation}

By incorporating the last WMA changing rate over the past 10 days, we capture the rate of change in visitor flows trends, providing our model with critical information about the direction and strength of recent trends.

\subsubsection{Event Features}

As discussed in Section \ref{sec:Social Media Popularity}, we establish three measures to assess event popularity on social media platforms: overall popularity, promotional popularity, and word-of-mouth popularity. We calculate these measures by event type, including music concerts, exhibitions, fireworks displays, and sports competitions. Notably, fireworks displays in this dataset are mostly single-session events held during major festivals in Hong Kong. Since the word-of-mouth effect does not apply to new or single-session events, we exclude word-of-mouth popularity for fireworks displays to adapt to this dataset.

To thoroughly evaluate the effectiveness of event popularity measures, we compare model performance across five distinct feature sets. As outlined in Table~\ref{tab:vars}, \textit{Feature Set 1} includes holidays, weather conditions, and temporal trends, while deliberately excluding event-related features. Building upon \textit{Feature Set 1}, \textit{Feature Set 2} incorporates the count of events by type, based on the assumption that the number of events held on a particular day could intuitively correlate with visitor numbers. To assess the impact of different popularity features, \textit{Feature Set 3} extends \textit{Feature Set 1} by adding overall popularity of events categorized by type, \textit{Feature Set 4} extends \textit{Feature Set 1} by including promotional popularity of events. Finally, \textit{Feature Set 5} integrates both word-of-mouth and promotional popularity features. By using these diverse feature sets, we aim to examine the comparative effectiveness of event popularity measures in predicting visitor flows.

\begin{table}
\caption{Description of key variables}\label{tab:vars}
\resizebox{\textwidth}{!}{%
\centering
\small  
    \begin{tabular}{p{2cm}|p{2.5cm}p{5.5cm}|p{1cm}p{1cm}p{1cm}p{1cm}p{1cm}}
     \hline
    Category & Features & Description & \textit{Feature Set 1}& \textit{Feature Set 2}& \textit{Feature Set 3}& \textit{Feature Set 4}&{Feature Set 5}
    \\
    \hline
        {Holiday 
        
        features} & {Holidays 
        
        remaining} & The number of holidays remaining according to the public holiday calendar of the tourists’ origin area (e.g., 3 means three holidays remain including today, and 0 represents a weekday). & \checkmark & \checkmark & \checkmark &\checkmark&\checkmark

        \\
        
        & Day before a holiday & A binary variable representing the day before a long public holiday (excluding weekends).  & \checkmark & \checkmark & \checkmark &\checkmark &\checkmark
        
        \\
        
        & {Week near a 
        
        holiday} & A binary variable representing the week before and after a long public holiday (excluding weekends).  & \checkmark & \checkmark & \checkmark &\checkmark &\checkmark
        \\
        
        & {Days to the 
        
        nearest holiday} & The number of working days to the last or next holiday (excluding weekends).   & \checkmark & \checkmark & \checkmark &\checkmark &\checkmark
        \\
        
        &School holidays  & A binary variable indicating the summer and winter vacations of most primary and middle schools in GBA during the study period.  & \checkmark & \checkmark & \checkmark &\checkmark &\checkmark

        \\
        \hline
        Weather &Rainfall  & The amount of daily total rainfall measured in millimeters. &  \checkmark & \checkmark & \checkmark &\checkmark &\checkmark
        \\
        
        &{Maximum 
        
        temperature }& The daily maximum temperature measured in Celsius. 
        & \checkmark & \checkmark & \checkmark &\checkmark &\checkmark
        \\
        
        &Typhoon & A binary variable representing days when a tropical cyclone warning signal is issued in Hong Kong. &   \checkmark & \checkmark & \checkmark &\checkmark &\checkmark
        \\
        \hline
        Temporal trend &Changing rate & The changing rate of the Moving Weighted Average of visitor flows over the past 10 days. &  \checkmark & \checkmark & \checkmark &\checkmark &\checkmark
        \\
        \hline
        {Event
        
        features} & Event count  & The numbers of music concerts, fireworks displays, exhibitions, and sports competitions & &  \checkmark &  &
        \\

        & {Overall 
        
        popularity}  &  The overall popularity of music concerts, fireworks displays, exhibitions, and sports competitions  & &  &\checkmark &
        \\
        
        & {Promotional 
        
        popularity} & The promotional popularity of music concerts, fireworks displays, exhibitions, and sports competitions &  &   & &\checkmark&\checkmark\\
        & {Word-of-mouth 
        
        popularity} & The word-of-mouth popularity of music concerts,  exhibitions, and sports competitions (fireworks displays are not included due to the scarcity of multi-session events). &  &  &   & &\checkmark
        \\

        \hline  
    \end{tabular}
    }
\end{table}

\subsection{Visitor Flows Prediction Model}

We develop a rolling prediction model based on GBDT to predict visitor flows. By incorporating a rolling window approach with a customized sample weighting mechanism, our model maintains adaptability and accuracy in the face of irregular demand dynamics during events \citep{LIU2023104725}. Originally popularized by \cite{friedman_greedy_2001}, GBDT constructs an ensemble of decision trees in a sequential fashion, with each new tree aimed at correcting the prediction errors of the existing ensemble. This method excels at capturing non-linear relationships and feature interactions, making it suitable for managing data characterized by irregular distributions and sparsity. These data characteristics are often present in event features, where time series values are frequently zero or missing, particularly concerning event popularity variables which exhibit sparsity and a long-tailed distribution. Thus, GBDT is particularly well-suited for capturing the effects of these features on visitor flows, providing robust predictions even in the face of complex data patterns.

To manage the irregular dynamics observed in visitor flows, we employ a rolling prediction method. This approach involves iterative training and testing in a rolling window, which mirrors real-world scenarios where operators continuously collect new data and predict visitor flows for subsequent days. Given a prediction horizon $\varphi$, the process begins by defining the training data from the initial time step $1$ to time $t$. We then predict values for the period extending from $t+1$ to $t+\varphi$. As the rolling prediction window slides by each time unit, the process is repeated, allowing the model to adapt to latest data points. For every time step, the model generates $\varphi$ predictions, subsequently averaging these predictions to provide a robust prediction. This approach accommodates the fluctuating dynamics of visitor flows as new information becomes available. To assess the robustness of this method in real-world operations, we compare rolling prediction horizons $\varphi$ ranging from 1 to 7 days. Long prediction horizons allows transportation authorities to gain sufficient time to collaborate closely with cross-city infrastructure operators, security personnel, emergency sectors, and event organizers to implement well-coordinated management plans.

Considering the tourism rebound in the post-COVID era, recent trends serve as more reliable indicators of future visitor flows. To leverage this, we incorporate a sample weighting mechanism into our predictive model. Specifically, we define a custom sample weight function that assigns greater weights to more recent data. This is achieved through the use of a sample weight decay factor, denoted as $ \delta $. The sample weight $ w_t $ for a given day $ t $ is determined by Eq.~\eqref{sampleweight}:

\begin{equation}
\label{sampleweight}
w_t = \max(1 - (T - t) \delta, 0)
\end{equation}
where $ T $ is the total length of training data.

To evaluate the performance of rolling predictions, we calculate the Mean Absolute Error (MAE) and the R-squared ($ R^2 $) metrics based on the average prediction for each iteration over the prediction horizon $\varphi$. Let $\hat{y}_{t}^{\text{avg}}$ represent the average prediction for time step $ t$ over all iterations:
\begin{equation}
\hat{y}_{t}^{\text{avg}} = \frac{1}{\varphi} \sum_{k=t-\varphi}^{t} \hat{y}_{k}
\end{equation}
where $\hat{y_k}$ represents the prediction for time step $k$. 

The MAE is calculated as:
\begin{equation}
\text{MAE} = \frac{1}{T} \sum_{t=1}^{T} \left| y_t - \hat{y}_{t}^{\text{avg}} \right|
\end{equation}
where $y_t$ represents the actual values at day $t$.

The $ R^2 $ score is defined by:
\begin{equation}
R^2 = 1 - \frac{\sum_{t=1}^{T} (y_t - \hat{y}_{t}^{\text{avg}})^2}{\sum_{t=1}^{T} (y_t - \bar{y})^2}
\end{equation}
where $\bar{y}$ is the mean of the actual values across all observations.

To evaluate the effectiveness of the GBDT rolling prediction model, we compare it against four commonly used prediction models as baselines: eXtreme Gradient Boosting (XGBoost), Random Forest (RF), Linear Regression (LR), and AutoRegressive Integrated Moving Average (ARIMA). We conduct comprehensive grid searches to identify the optimal hyperparameters for each model. Additional model details are provided in ~\ref{modeldetail}.

\section{Results}
\label{sec1}

\subsection{Feature Sets Assessment and Model Comparison}

After events filtration, we compile an event dataset comprising 369 events and 1,112 sub-sessions, spanning over 14 months from March 1, 2023, to May 15, 2024. To predict cross-border visitor flows from mainland China, we employ a rolling prediction model utilizing all data available up to each predictive step. We conduct predictions using five models across various feature sets over a one-day horizon. Table~\ref{tab:pop_select} details the performance of each model and feature set combination. Notably, the GBDT model with \textit{Feature Set 5} achieves the highest accuracy, with an $R^2$ of 85.71\%.

Incorporating promotional popularity and word-of-mouth popularity features, \textit{Feature Set 5} enhances the $R^2$ by more than 5\% compared to the same model without event features (\textit{Feature Set 1}). While \textit{Feature Set 2} (with event counts) provides some improvement compared to \textit{Feature Set 1}, it falls short compared to other sets with event popularity features. \textit{Feature Set 3} does not distinguish the timing of social media posts, relying purely on overall popularity, whereas \textit{Feature Set 4} focuses only on promotional popularity, excluding word-of-mouth effects. The predictive performance of \textit{Feature Sets 3 and 4} is inferior to \textit{Feature Set 5}, underlining the importance of accounting for both promotional and word-of-mouth effects on social media for accurate visitor flows prediction. Similarly, in the XGBoost and RF models, the best predictive accuracy is also observed with \textit{Feature Set 5}. In contrast, the LR and ARIMA models perform best with \textit{Feature Set 2}. This discrepancy is due to the LR and ARIMA models' limitations in effectively processing the sparsely and irregularly distributed event popularity variables.

\begin{table} 
\caption{Model results with different feature sets}\label{tab:pop_select}
\centering\footnotesize
\begin{tabular}{p{5.5cm}|p{1cm}p{1.2cm}p{1.5cm}p{1.2cm}p{1.5cm}p{1.2cm}}
    \toprule
    Feature Combinations & & GBDT & XGBoost & RF & LR &ARIMA\\ 
    \midrule
    {\textit{Feature Set 1}: 
    
    } & $R^2$ &80.42 & 80.69 &79.63&62.69 &67.10\\
    Without event features& $MAE$ & 9499.59 & 9388.53 & 9717.39 & 12291.99 &11162.33 \\
    \midrule
    {\textit{Feature Set 2}:

    }  &$R^2$& 83.91 & 84.10 & 80.30 & \underline{64.59} &  \underline{70.94}\\
    With event count  & $MAE$ & 9078.23& 8780.26 & 9713.02 &\underline{12243.88} & \underline{10760.19}\\
    \midrule
    
    \textit{Feature Set 3}: 
    
 &$R^2$& 84.69 & 84.10 & 80.41 &36.17 &  37.44\\
        
        With overall popularity  
    & $MAE$ & 8783.74& 9604.99 & 9641.75 &12950.05 &12153.52\\
    \midrule
    
    {\textit{Feature Set 4}:

    } &$R^2$&   84.57  & 84.03 & \underline{80.48} &59.45&68.69\\
    
    With promotional popularity & $MAE$ & 8905.10 & 8779.05 & 9642.43 &12560.84 &11286.73\\
    \midrule
    
    {\textit{Feature Set 5}: 
    
    With promotional popularity }  & $R^2$ & \underline {85.71} & \underline{84.83} & \underline{80.48} &59.08&68.19 \\
    
    {and word-of-mouth popularity}& $MAE$ & \underline{8549.37} & \underline{8681.38} & \underline{9586.91}  &  12906.80 &11516.99
    \\
    \bottomrule
\end{tabular}
\end{table}

\subsection{Event-aware Visitor Flows Prediction}

\begin{figure}[!ht]
  \centering
  \includegraphics[width=1\textwidth]{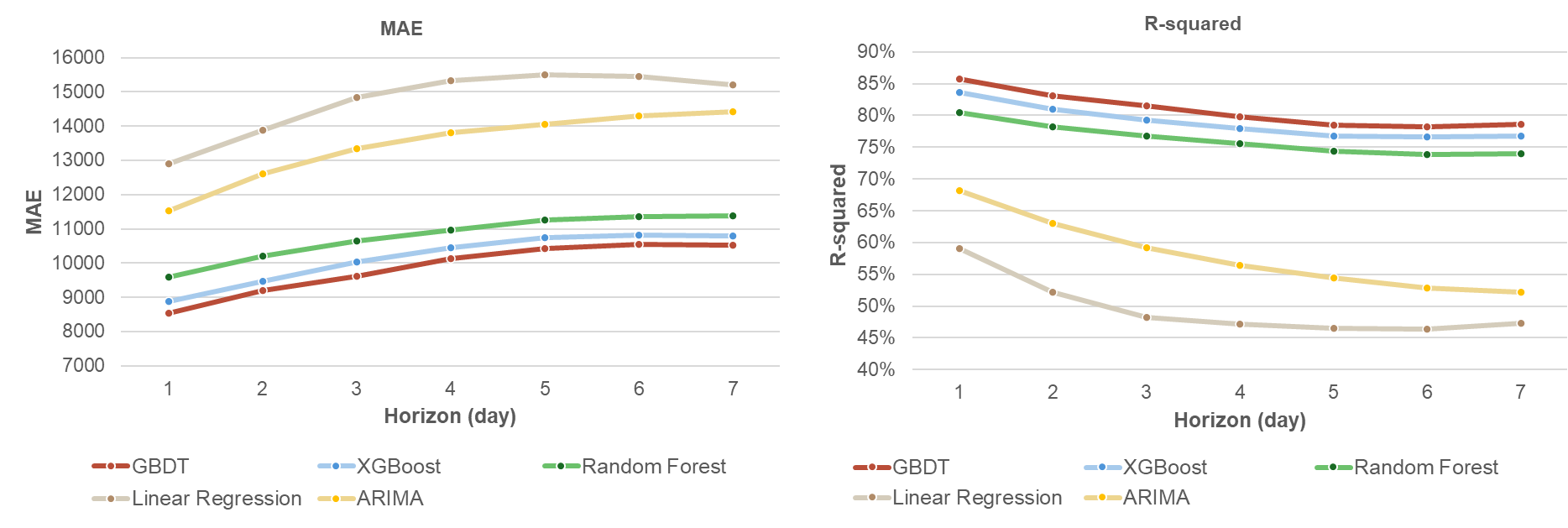}
  \caption{Model comparison of prediction accuracy under different horizons}\label{fig:horizon}
\end{figure}

\begin{figure}[!ht]
  \centering
  \includegraphics[width=1\textwidth]{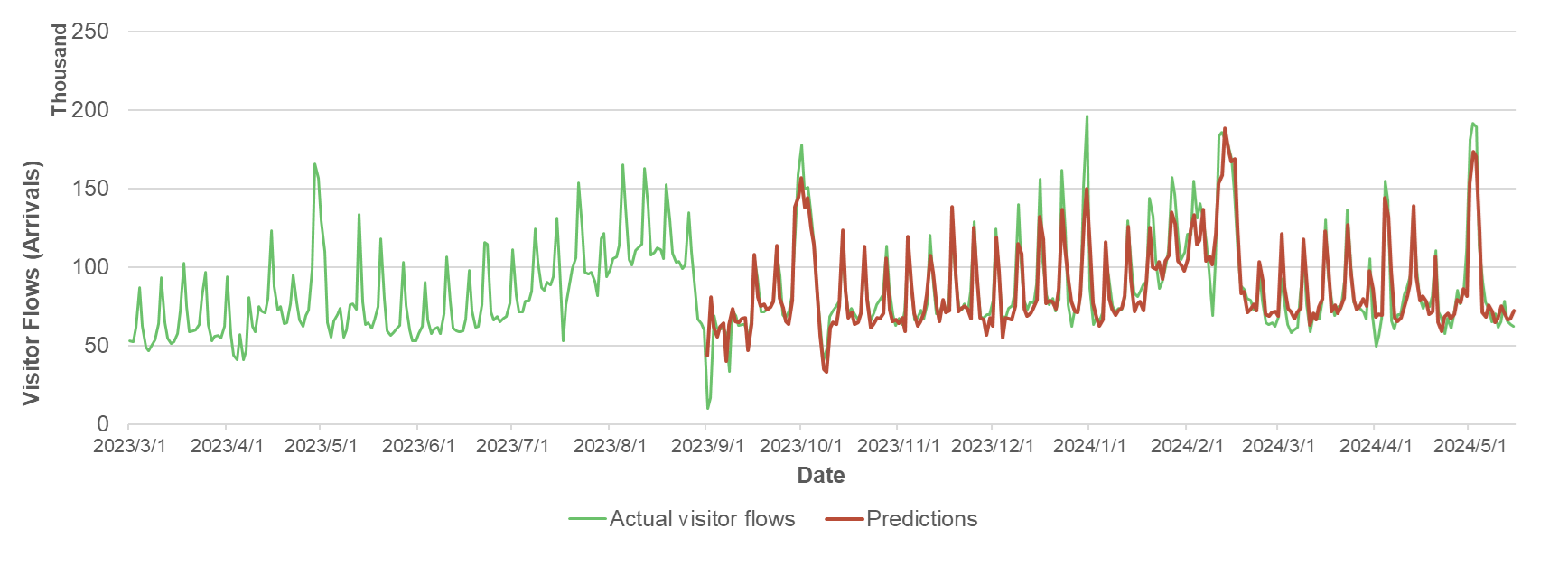}
  \caption{GBDT rolling prediction on visitor flows under one-day horizon}\label{fig:prediciton}
\end{figure}

To test the model's robustness in real-world applications, we predict cross-border visitor flows over horizons ranging from 1 to 7 days, which emulates how operators gradually incorporate newly available data points in prediction and make arrangements proactively. While longer horizons provide more time for operational arrangements, they typically yield lower accuracy. Fig.~\ref{fig:horizon} illustrates the performance of the five models across different prediction horizons with \textit{Feature Set 5}. The GBDT model, represented in red, outperforms the other models, exhibiting the lowest MAE (8549 $\sim$ 10512) and the highest $R^2$ scores (85.71\% $\sim$ 78.62\%). It is followed by the XGBoost and Random Forest model. The ARIMA and Linear Regression models significantly underperform compared to the ensemble learning methods. Fig.~\ref{fig:prediciton} illustrates the predictions generated by GBDT model over a one-day horizon, where the predicted visitor flows are depicted in red and the actual visitor flow flow is represented in green. Although predicting irregular travel demand surges is always challenging without sufficient historical data, our event-aware model achieves a reasonable accuracy.

\subsection{Factors Affecting Visitor Flows}

\begin{figure}[!ht]
  \centering
  \includegraphics[width=0.6\textwidth]{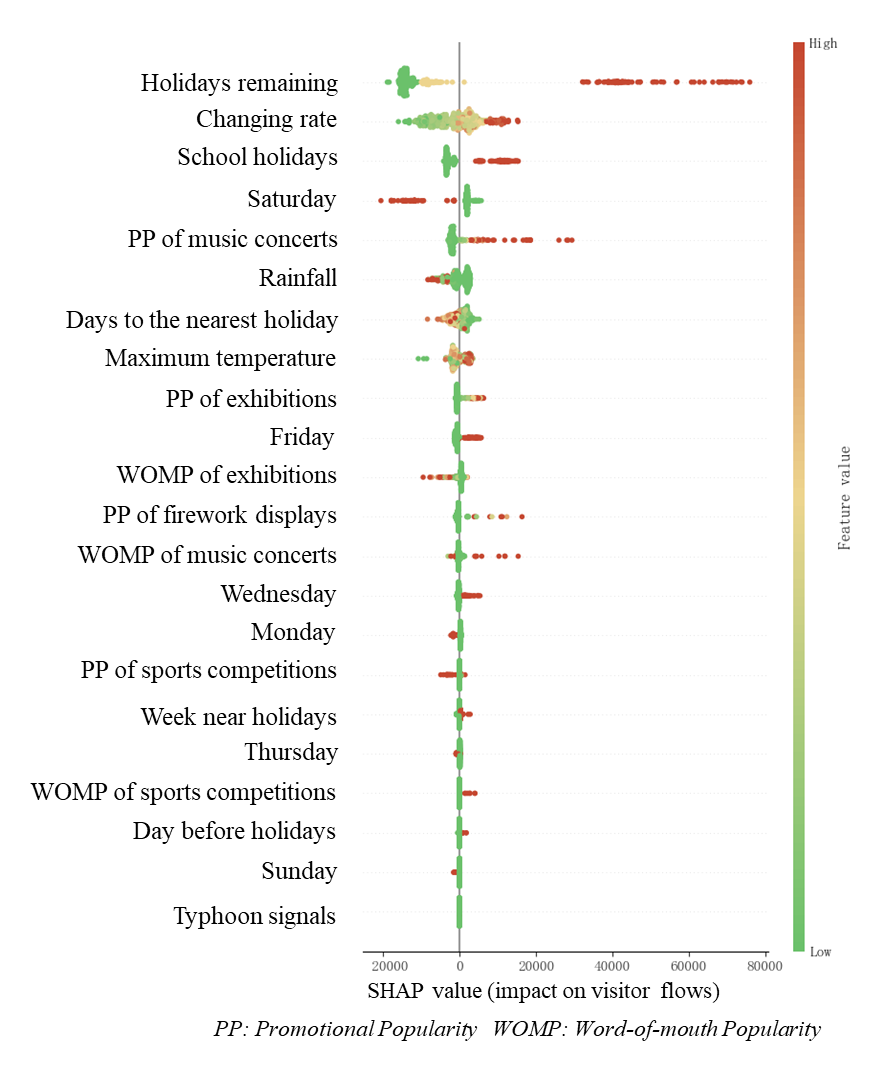}
  \caption{Feature importance on visitors flow}\label{fig:Shap}
\end{figure}

Understanding the factors influencing visitor flows is crucial for developing effective operational strategies for intercity transport infrastructure and services. We employ Shapley Additive Explanations (SHAP) values to explain the importance and effects of the different features for the best model (GBDT under a one-day horizon with \textit{Feature Set 5}). 
SHAP values interpret each feature's impact on cross-border visitor flows. In Fig.~\ref{fig:Shap}, the y-axis ranks features by descending importance, and the x-axis shows their impact on visitor flows. Features with clusters extending far from the center line have stronger impacts, with positive impacts to the right and negative impacts to the left.

Among the features, the number of days remaining in the holiday is the primary driving factor affecting cross-border visitor flows, with a large number of visitors arriving at the beginning of a holiday. The WMA changing rate, which indicates recent trends, also plays a crucial role in visitor flows prediction. School holidays, such as summer and winter breaks, typically correspond to the peak tourism seasons, which are associated with a high number of visitor flows. Adverse weather conditions, such as heavy rain, negatively impact cross-border visitor flows. The maximum temperature also exhibits high importance in the prediction, suggesting the seasonal fluctuation of visitor flows.

Visitors are also largely motivated by an event's popularity on social media platforms to embark on cross-city trips. The promotional popularity of music concerts, exhibitions, and fireworks displays positively impacts cross-border visitor flows. Among these events, increased promotional popularity for exhibitions has a smaller effect on attracting visitors compared to fireworks displays and music concerts. This variation arises because the extent to which social media engagement materializes into actual visitor flow numbers can differ across event types.
It suggests that a smaller percentage of users who interact with promotional posts for exhibitions actually attend, in contrast to those engaging with posts about concerts and fireworks displays. In addition, exhibitions are often held over a long period. Although we excluded events with more than 30 sessions during data processing, the remaining exhibitions still average six sessions, significantly more than the average of three sessions for other event types. The multi-session nature of exhibitions causes attendees to be distributed across many sessions, diluting the impact on daily visitor flows.

The findings provide valuable insights into the effect of word-of-mouth on different types of events. For music concerts, an increase in word-of-mouth popularity positively impacts visitor flows. Interestingly, for exhibitions, a rise in word-of-mouth popularity seems to have a negative effect on visitor numbers. This is because exhibitions typically run for extended periods, with visitor numbers often peaking during the opening days and then gradually declining to a low but stable level. The word-of-mouth popularity on social media may not be sufficient to counteract this downward trend in attendance over time. We conducted further analysis to validate this interpretation, with detailed findings presented in ~\ref{exhibitiondetail}.


In addition, the promotional and word-of-mouth popularity features of sports competitions are less important in predicting visitor flows. This indicates that the social media exposure of sports competitions in Hong Kong may not significantly attract visitors from mainland China. This could be because these events are not as competitive as the high-profile international sports competitions held frequently in mainland China.

\subsection{Events' Impact Across Travel Modes}

To demonstrate the practical relevance of our model for the efficient operation of cross-city infrastructure, we apply it to analyze mainland Chinese visitor flows at three of Hong Kong's busiest border entry points. These entry points, each offering distinct transport services, are the Lok Ma Chau Spur Line (for metro service), West Kowloon HSR Station, and Hong Kong International Airport. Our findings reveal that the impact of events on visitor flows varies across these entry points. Visitors' choices of entry points are closely tied to their trip origins, as each mode of transportation caters to different travel distances. This suggests that events have varying levels of appeal for visitors from different areas. Specifically, the metro entry point primarily facilitates travel between Hong Kong and Shenzhen, a neighboring city. In contrast, travelers arriving at the HSR entry point are mainly from cities within the GBA and nearby regions. Airport arrivals typically originate from more distant cities.

\begin{figure}[!ht]
  \centering
  \includegraphics[width=1\textwidth]{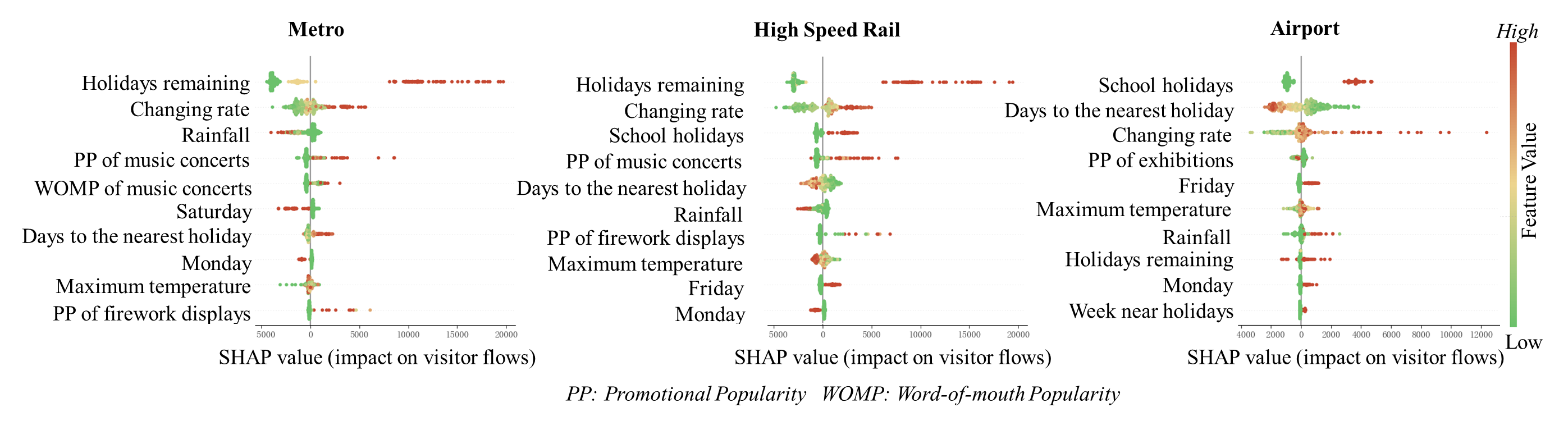}
  \caption{Feature importance on visitors flow by border entry points (Top 10 features)}\label{fig:Shap_borders}
\end{figure}

Figure \ref{fig:Shap_borders} illustrates the top ten features affecting visitor flows at the three entry points. Visitor flows at both the metro and HSR entry points are positively associated with the promotional popularity of events such as concerts, fireworks displays. Visitors arriving by metro, primarily from Shenzhen, exhibit high travel flexibility and are more responsive to the word-of-mouth effect of concerts. They are more inclined to attend subsequent concerts after noticing their rising popularity on social media. In contrast, HSR travelers, often from cities farther away, tend to plan their trips in advance and are less swayed by word-of-mouth of events to make spontaneous cross-city trips. Compared to visitors taking metro, those from cities connected by HSR are more responsive to online fireworks displays promotions. The distance decay effect suggests that travel demand diminishes as geographical distance increases. 
For visitors taking long-haul flights, their trips are typically planned well in advance and are less flexible. The number of visitors by air is more closely linked to holidays rather than events. 
The analysis reveals the heterogeneous impacts of events on visitor flows via different transport modes, providing valuable insights for event-aware intercity transport infrastructure planning and management strategies.

\section{Conclusion and Discussion}


\subsection{Key Findings and Contributions}

This study makes a significant contribution by developing a novel LLM-based pipeline to compile comprehensive event datasets at city-level, covering diverse and concurrent events, which addresses a notable gap in existing research. 
Leveraging the robust text-processing capabilities of LLMs, the pipeline efficiently extracts event features from online textual data and enables semantic analysis of social media data. To our knowledge, this study is the first to utilize large-scale social media data to estimate the potential impacts of events on cross-city visitor flows. 

Furthermore, this study defines specific metrics to capture the promotional and word-of-mouth effects on social media platforms. These effects influence people's intentions to undertake cross-city trips to attend events and can therefore be used to estimate an event's online popularity \citep{hu_tourism_2022,cui_operational_2018,duan_online_2008,dellarocas_exploring_2007}. By operationalizing these effects into measurable indicators, the study proposes an event-aware machine learning model that enhances the performance of visitor flow prediction. Through the case study of Hong Kong, we further reveal the heterogeneous effects across event types and identify key travel modes that are susceptible to the disruptions caused by these events. Importantly, the generalizability of this approach allows its application in other regions experiencing similar event-motivated cross-city travel dynamics.

\subsection{Practical Implications for Transport Management}

This study offers significant practical implications for managing transport systems under diverse and concurrent events. Effective management requires coordinated policy measures to ensure efficient transport operations at multiple levels. Timely and localized information on predicted visitor flows and the potential impacts of events should be shared among relevant parties, including cross-city facility managers, public transit providers, parking facility managers, traffic police and emergency service agencies, and event organizers. By disseminating this information, stakeholders can collaboratively develop strategies to address the challenges posed by event-driven travel demand, enhancing the resilience and efficiency of transportation systems.

For cross-city travel, integrating event features and social media analytics into visitor flow prediction models significantly improves prediction performance and helps explain seemingly irregular demand fluctuations. Since cross-city transport infrastructure and services are primarily designed for regular travel demand, sudden surges in visitor flows often lead to overcrowding and congestion. Incorporating event-awareness into regional travel demand forecasting can inform future planning and development of cross-city transport infrastructure. For the management of existing transport infrastructure, transport policymakers can leverage these insights to adjust resource allocation and operational strategies at major entry points.

In cross-border contexts such as Hong Kong, event-aware visitor flow prediction across borders can help identify those entry points susceptible to disruptions caused by events. Such insights enable border control authorities to enhance processing efficiency at key entry points and minimize delays through adaptive resource allocation. Proactive adjustments, such as deploying additional customs personnel, extending service hours, or opening extra processing lanes, can help accommodate anticipated demand surges, ensuring smoother cross-border mobility.

When the expected demand surges exceed a certain threshold, local public transit systems also need to be adapted to better serve trips to event venues. With the proposed rolling prediction method, public transit operators can continuously refine and adjust operational strategies to scale up services and prevent overcrowding. Introducing special shuttle services between major cross-city transport hubs and event venues, such as temporary night shuttles connecting HSR stations to event sites, can enhance connectivity. Additionally, public transit should increase service frequency and extend operating hours before and after large-scale events. To further balance transport demand during peak periods, on-demand mobility services can play a complementary role by implementing strategies such as expanding bike-sharing availability around event sites or designating dedicated ride-hailing pick-up or drop-off points.

This study highlights the varied impact of online popularity across event types, allowing us to identify events with significant online popularity and potential to attract large crowds. For these high-impact events, comprehensive on-site traffic and crowd management strategies are necessary. Extending prediction horizons allows transportation authorities to gain sufficient lead time to collaborate closely with event organizers, parking facility managers, traffic police and emergency service agencies to implement well-coordinated management plans. These measures may include establishing temporary pedestrian zones, providing crowd navigation information, implementing traffic re-routing strategies, optimizing parking management, and enhancing surveillance systems. Collectively, these efforts aim to ensure public safety and maintain efficient mobility around the event venue.

\subsection{Broader Strategic Implications}

Organizing large-scale public events in cities requires collaboration across multiple sectors. These analytical results can help foster cross-departmental coordination and amplify the collective economic benefits. Understanding the impact of various events on visitor flows is essential not only for transport management but also for broader event planning, tourism development, and destination branding strategies. In the case of Hong Kong, SHAP analysis demonstrates that events such as music concerts, fireworks displays, and exhibitions with high promotional popularity are closely associated with increased visitor demand. Destination branding and tourism authorities can optimize event planning strategies to prioritize events with greater visitor appeal or coordinate event portfolios \citep{ziakas_event_2021,ziakas_leveraging_2023}. Event portfolio management allows concurrent events to generate a synergistic effect, providing a richer and more diverse array of attractions that appeal to wider visitor groups. 
Additionally, the hospitality and retail sectors can leverage these predictive insights to make informed decisions regarding inventory management, staffing, and dynamic pricing strategies. By anticipating surges in visitor demand during high-profile events, businesses can optimize operations, enhance customer experience, and maximize profitability \citep{wu_topic_2024}. 
Considering the potential origins of visitors using different travel modes, the study highlights the varied geographical impact of pre-event promotions and post-event word-of-mouth communication. These insights help event and destination marketing organizations optimize the allocation of online marketing resources, enabling them to effectively target key audiences, enhance advertising efficiency, and maximize revenue.

\subsection{Limitations and Future Research}

While this study offers valuable insights, it does have certain limitations. First, although we define event popularity through social media engagement, it is important to note that users who ``like" or ``collect" posts do not necessarily plan to attend the events \citep{onder_utilizing_2020}. Even among those expressing the intent to do so, there is often a significant gap between intention and actual attendance behavior \citep{lee_estimating_2014}. Therefore, in our study, social media popularity primarily reflects an event's exposure rather than serving as a reliable measure of actual attendance. Second, our analysis of social media popularity mainly utilizes the platform \textit{Rednote}, a prevalent Chinese social media network. Although the platform's main user demographic aligns with key groups of event visitors and cross-border visitors, \textit{Rednote} has a higher proportion of young and female users. This may slightly skew the representation of the true demographic distribution of event visitors. To mitigate this bias, future research may apply the proposed framework to multiple social media platforms and compare the results. 

Looking ahead, future research could delve into the economic impact of cross-city mobility motivated by diverse events. Building on the findings regarding the effectiveness of different events and their online popularity, researchers can further evaluate the overall economic benefits of hosting different types of events.
Moreover, future studies could incorporate individual mobility data, such as smart card usage or mobile phone traces, to better understand visitors' spatiotemporal mobility patterns during events. This data would enable a more detailed analysis of event-motivated human mobility within a city. Also, individual mobility data can be used to estimate the temporal flow of visitors around specific event venues. This allows for a more nuanced understanding of how social media popularity is related to spatiotemporal travel demand patterns throughout the event's entire lifecycle.

\section*{Acknowledgment}
This research is supported by National Natural Science Foundation of China (NSFC 42201502).

\appendix
\section{Model Details}
\label{modeldetail}
GBDT builds an additive model in a forward stage-wise manner. It optimizes an objective function that consists of a differentiable loss function and a regularization term. Let $\mathcal{L}(y_t, F(x_t))$ denote the loss function where $ y_t $ is the true value and $ F(x_t) $ is the prediction. Here, $ x_t $ represents the feature vector for day $ t $. The objective function $J$ to be defined as:

\begin{equation}
 J =  min_J \sum_{t=1}^T w_t \mathcal{L}(y_t, F_{i-1}(x_t) + \gamma h_i(x_t)) + \Omega(h_i)
\end{equation}
where $ w_t $ denotes the sample weight for day $ t $, $ T $ the total length of training data (number of days). $ F_{i-1}(x_t) $ represents the prediction from the ensemble of trees up to the $(i-1)^{\text{th}}$ iteration, $ h_i(x_t) $ represents the prediction from the $ i^{\text{th}} $ new tree.
$ \gamma $ is the learning rate, $ \Omega(h_i) $ is the regularization term to penalize the complexity of the model.

We perform a comprehensive grid search to determine the optimal hyperparameters. As detailed in Table~\ref{tab2}, we systematically adjust the learning rate ($\gamma$), maximum depth ($d$), and the number of estimators (trees) ($I$). Additionally, we vary the sample weight decay factor ($\delta$) within the loss function. This grid search identifies those hyperparameters that result in the best performance as measured by the $R^2$ value. In the GBDT model, the hyperparameters are optimized as follows: a learning rate of 0.05, a maximum depth of 3, 500 estimators, and a sample weight decay factor of 0.005. 

\begin{table}[ht!] 
\caption{Grid Search for Optimal Parameters}\label{tab2}
\centering
\begin{tabular}{l| c}
\hline
    Parameters & Steps  \\
    \hline
    Learning rate $ \gamma $ & 0.01, 0.05, 0.1  \\
    Maximum depth $ d $ & 3, 5, 7  \\
    Number of estimators $ I $ & 100, 200, 500, 1000  \\
    Sample weight decay factor $\delta$ & 0, 0.001, 0.002, 0.003, 0.004, 0.005, 0.006  \\
    \hline
\end{tabular}
\end{table}

We compare the GBDT model against four commonly used baseline models: XGBoost, RF, LR, and ARIMA. 
To ensure optimal performance for each baseline model, we apply a grid search on the XGB and RF models to determine the best hyperparameters, including learning rate, maximum depth, number of estimators, and sample weight decay factors. For the ARIMA model, we use a grid search to optimize its specific parameters: autoregression order $p$, degree of differencing $d$, and moving average order $q$. In the case of the LR model, we focus on optimizing its sample weight decay factors. We also implement the same rolling prediction mechanism to the XGB, RF, and LR models.

\section{Further Analysis of Exhibitions' Word-of-mouth Popularity}
\label{exhibitiondetail}

\begin{figure}[!ht]
  \centering
  \includegraphics[width=0.6\textwidth]{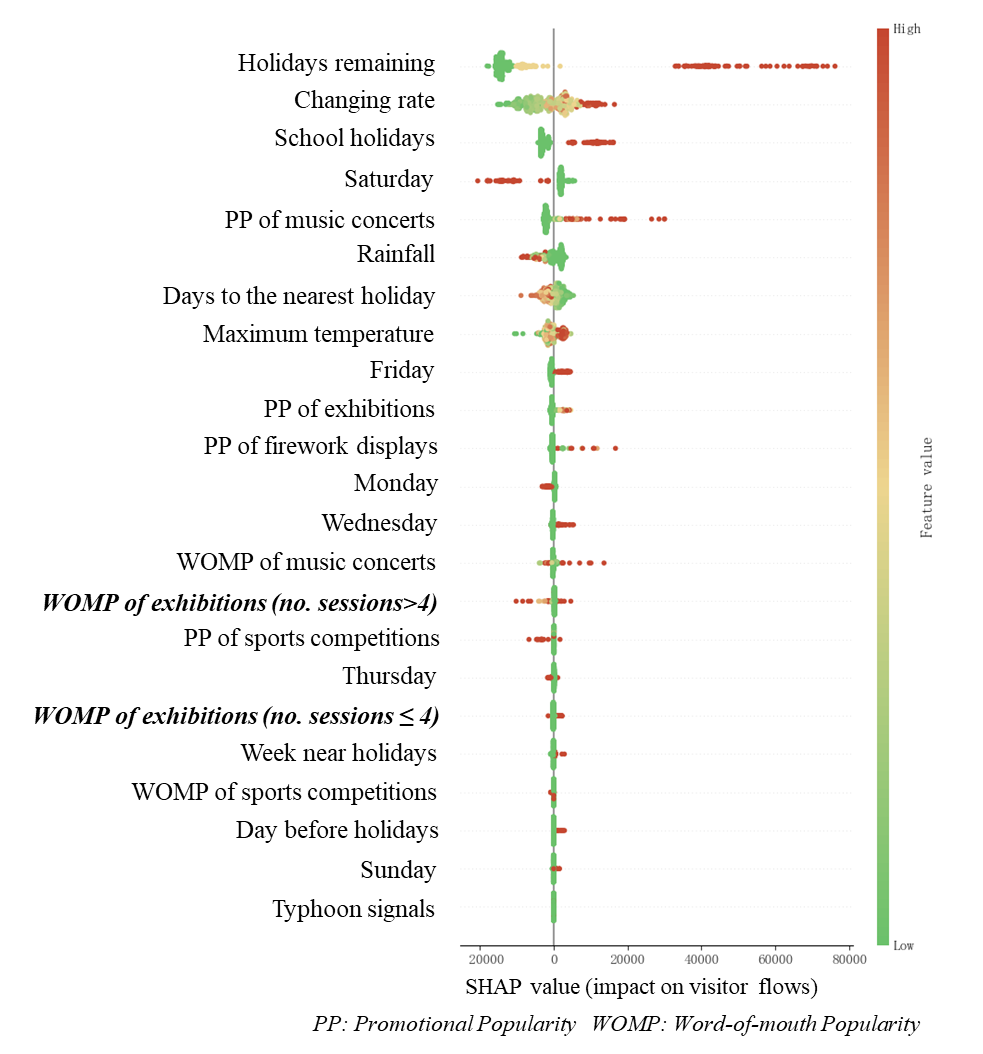}
  \caption{Feature interpretation on exhibitions' word-of-mouth popularity }\label{fig:Shap_exhibition}
\end{figure}

To further explore the counterintuitive correlation between the popularity of word-of-mouth of exhibitions and visitor flow, we replaced the general ``word-of-mouth popularity of exhibitions'' with two specific variables that account for the exhibition sessions. The first variable measures the word-of-mouth popularity during the initial four days of the exhibition, while the second assesses it in the subsequent days. As shown in Fig.~\ref{fig:Shap_exhibition}, higher word-of-mouth popularity during the first four days correlates with increased visitor numbers. This indicates that the positive effects of word-of-mouth are most pronounced before the natural decline in attendance begins. However, after the fourth day, the relationship between word-of-mouth popularity and visitor numbers becomes inconsistent, showing no clear pattern. To gain a deeper understanding of the role of word-of-mouth and the temporal decline in attendance across the exhibition's lifecycle, future studies would benefit from more detailed data on attendance trends over time.

\bibliographystyle{model5-names2}\biboptions{authoryear}
\bibliography{main}

\end{document}